\documentclass[preprint,floats,epsfig]{revtex4}
\usepackage{epsf,epsfig}
\usepackage{amsmath}
\usepackage{subfigure}
\usepackage{graphicx}
\usepackage{feynmf}

\preprint{
\hbox to \hsize{
\hfill$\vcenter{\hbox{\bf MADPH-07-1498}
                \hbox{\bf hep-ph/0710.4147}
                \hbox{October 2007}
		\hbox{}
		\hbox{}}$
		}
}

\begin{document}

%-----------------------------------
% Title
%-----------------------------------
\title{\vspace*{.75in}
Cosmological and Astrophysical Constraints \\
on Tensor Unparticles}

%-----------------------------------
% Authors
%-----------------------------------
\author{
Ian Lewis
}

%-----------------------------------
% Address
%-----------------------------------
\affiliation{
Department of Physics, University of Wisconsin, 1150 University Avenue, Madison, WI 53706 
\vspace*{.5in}}

\begin{abstract}
\noindent
We calculate cosmological and astrophysical bounds on the couplings between standard model fields and tensor unparticles.  The present day density of tensor unparticles from neutrino-neutrino and photon-photon annihilation is calculated.  Also, the supernovae volume energy loss rates from electron-positron and photon-photon annihilation to tensor unparticles are calculated.  The constraints from matter density and supernovae volume energy loss rates from photon-photon annihilation are on the same order of magnitude, while the bounds from supernovae volume energy loss rates from electron-positron annihilation are an order of magnitude lower.  We find the couplings between standard model fields and tensor unparticles are at least an order of magnitude lower than those used for previous studies of tensor unparticle collider phenomenology.
\end{abstract}
\pacs{}

\maketitle
\section{Introduction}
Recently Georgi \cite{Georgi:2007ek,Georgi:2007si} has introduced the idea of a low energy theory where a scale-invariant sector couples to the standard model.  At high energies a theory with a non-trivial infrared fixed point couples to the standard model via particles with mass $M_{\mathcal{U}}$.  Below $M_{\mathcal{U}}$ the non-renormalizable interaction is

\begin{eqnarray}
\frac{1}{M^{k}_{\mathcal{U}}}O_{SM}O_{\mathcal{BZ}},
\label{eqn:BZOP}
\end{eqnarray}
where $k=n+d_{\mathcal{BZ}}-4$, $n$ is the scale dimension of the standard model operator $O_{SM}$, and $d_{\mathcal{BZ}}$ is the scale dimension of the operator $O_{\mathcal{BZ}}$, labeled $\mathcal{BZ}$ for Banks-Zaks \cite{Banks:1981nn}.  Since the mediating particles have not been observed, we expect their masses, $M_{\mathcal{U}}$, to be larger than $~1$ TeV.

\par
  The scale invariance of the Banks-Zaks operator sets in at a scale $\Lambda_{\mathcal{U}}$.  At energies below $\Lambda_{\mathcal{U}}$, 
the conformal sector couples to the standard model through the non-renormalizable interactions

\begin{eqnarray}
C_{\mathcal{U}}\frac{\Lambda^{d_{\mathcal{BZ}}-d_{\mathcal{U}}}_{\mathcal{U}}}{M^{k}_{\mathcal{U}}}O_{SM}O_{\mathcal{U}}\equiv\frac{\kappa_{\mathcal{U}}}{\Lambda^{d_{\mathcal{U}}+n-4}_{\mathcal{U}}}O_{SM}O_{\mathcal{U}},~~\kappa_{\mathcal{U}}=C_{\mathcal{U}}\bigg{(}\frac{\Lambda_{\mathcal{U}}}{M_{\mathcal{U}}}\bigg{)}^k
\label{eqn:UNPARTOP}
\end{eqnarray}
where $d_{\mathcal{U}}$ is the dimension of the unparticle operator $O_{\mathcal{U}}$, and $C_{\mathcal{U}}$ are dimensionless constants set by matching Eqs.~(\ref{eqn:BZOP}) and (\ref{eqn:UNPARTOP}) below the conformal scale $\Lambda_{\mathcal{U}}$.  In principle $C_{\mathcal{U}}$ can be different for each standard model operator, but we expect them to be on the order of $1$. For the unparticles to be relevent at future colliders, such as the LHC, $\Lambda_{\mathcal{U}}$ should be at least a few $100$ GeV.

\par
If the unparticle sector is to be scale invariant, its fields cannot have definite non-zero mass.  Hence, their spectral density function is different than that of ordinary particles.  Demanding that the matrix element $\langle 0|O_{\mathcal{U}}(x)O^\dagger_{\mathcal{U}}(0)|0\rangle$ be scale invariant and have scaling dimension $2d_{\mathcal{U}}$, the spectral density function of unparticles is found to be \cite{Georgi:2007ek}

\begin{eqnarray}
\rho(P^2_{\mathcal{U}})=A_{d_{\mathcal{U}}}\theta(P^0_{\mathcal{U}})\theta(P^2_{\mathcal{U}})(P^2_{\mathcal{U}})^{d_{\mathcal{U}}-2},
\end{eqnarray}
where, by convention,

\begin{eqnarray}
A_{d_{\mathcal{U}}}=\frac{16\pi^{5/2}}{(2\pi)^{2d_{\mathcal{U}}}}\frac{\Gamma(d_{\mathcal{U}}+1/2)}{\Gamma(d_{\mathcal{U}}-2)\Gamma(2d_{\mathcal{U}})}.
\end{eqnarray}

There has been much recent interest in the phenomenological \cite{Cheung:2007ue, Luo:2007bq, Chen:2007vv,Ding:2007bm,Liao:2007bx,Aliev:2007qw,Li:2007by,Duraisamy:2007aw,Lu:2007mx,Fox:2007sy,Greiner:2007hr,Choudhury:2007js, Chen:2007qr,Aliev:2007gr,Mathews:2007hr,Zhou:2007zq,Ding:2007zw,Chen:2007je,Bander:2007nd,Rizzo:2007xr,Cheung:2007ap,Chen:2007zy,Zwicky:2007vv,
Kikuchi:2007qd,Mohanta:2007ad,Huang:2007ax,Lenz:2007nj,Choudhury:2007cq,Zhang:2007ih,Li:2007kj,Deshpande:2007jy,Mohanta:2007uu,
Cacciapaglia:2007jq,Neubert:2007kh,Luo:2007me,Bhattacharyya:2007pi,Majumdar:2007mp,Alan:2007ss,Chen:2007pu,Hur:2007cr,
Anchordoqui:2007dp,Balantekin:2007eg,Aliev:2007rm,Iltan:2007ve,Chen:2007cz,Majhi:2007tu,Kumar:2007af,Ding:2007jr,Kobakhidze:2007zs}, astrophysical \cite{Hannestad:2007ys,Das:2007nu,Freitas:2007ip}, long-range force \cite{Liao:2007ic,Deshpande:2007mf,Das:2007cc,Freitas:2007ip}, and cosmological \cite{Davoudiasl:2007jr,McDonald:2007bt} implications of unparticle physics.

\par
In this paper we look at the limits on the couplings between standard model fields and tensor unparticles from cosmology and astrophysics.  In Section \ref{sec:Model} we present effective interactions between tensor unparticles and standard model fields and outline the deconstruction of unparticles.  The present day tensor unparticle density is calculated and matter density constraints are placed on the tensor unparticle and standard model couplings in Section \ref{sec:density}.  In Section \ref{sec:emissivity}, the supernovae volume energy loss rates from photon-photon and electron-positron annihilation to tensor unparticles is calculated and constraints from SN1987A are placed on the tensor unparticle and standard model couplings.  A summary of results and the conclusions are given in Section \ref{sec:Conclusion}.

\section{The Model}
\label{sec:Model}
The effective standard model and tensor unparticle interactions under study are \cite{Cheung:2007ap}
\begin{eqnarray}
\frac{\kappa_\gamma}{\Lambda^{d_{\mathcal{U}}}_{\mathcal{U}}}F_{\mu\beta}F^{\beta}_{\nu} O^{\mu\nu}_{\mathcal{U}},~~~-\frac{1}{4}\frac{\kappa_f}{\Lambda^{d_{\mathcal{U}}}_{\mathcal{U}}}\overline{\psi}_fi(\gamma_\mu  \stackrel{\leftrightarrow}{D}_\nu+\gamma_\nu \stackrel{\leftrightarrow}{D}_\mu)\psi_f O^{\mu\nu}_{\mathcal{U}},\label{eq:UNPARTINT}
\end{eqnarray}
where $F^{\mu\nu}$ is the photon field strength operator, $\psi_f$ is a fermion field, and $O^{\mu\nu}_{\mathcal{U}}$ is the tensor unparticle operator.

\par
Following the method in \cite{Stephanov:2007ry}, we break the conformal invariance of the unparticle by decomposing it into fields $\lambda^{\mu\nu}_j$ with masses
\begin{eqnarray}
M^2_j=\Delta^2j,
\label{eq:massgap}
\end{eqnarray}
where $\Delta$ is some mass gap.  As $\Delta\rightarrow 0$, the mass spectrum of the $\lambda_j$'s becomes continuous.  This limit is the conformal limit.
\par
 Now we rewrite the unparticle operator as $O^{\mu\nu}_{\mathcal{U}}=\sum_jF_j\lambda^{\mu\nu}_j$.  By matching the correlation functions of the scale-invariant unparticle to that of the deconstructed unparticle in the conformal limit, it can be found that 
\begin{eqnarray}
F^2_j=\frac{A_{d_{\mathcal{U}}}}{2\pi}\Delta^2(M^2_j)^{d_{\mathcal{U}}-2}.
\end{eqnarray}
The interactions in Eq. (\ref{eq:UNPARTINT}) then become

\begin{eqnarray}
\frac{\kappa_\gamma}{\Lambda^{d_{\mathcal{U}}}_{\mathcal{U}}}F_{\mu\beta}F^{\beta}_{\nu}\sum_jF_j\lambda^{\mu\nu}_j,~~~-\frac{1}{4}\frac{\kappa_f}{\Lambda^{d_{\mathcal{U}}}_{\mathcal{U}}}\overline{\psi}_fi(\gamma_\mu  \stackrel{\leftrightarrow}{D}_\nu+\gamma_\nu \stackrel{\leftrightarrow}{D}_\mu)\psi_f\sum_jF_j\lambda^{\mu\nu}_j.\label{eq:DECOMPINT}
\end{eqnarray}
Hence, the Feynman rules for the fields $\lambda^{\mu\nu}_j$ are the same as those found in \cite{Cheung:2007ap} for $O^{\mu\nu}_{\mathcal{U}}$ multiplied by $F_j$.

\section{Critical Density Bounds}
\label{sec:density}

Since $F_j\rightarrow 0$ in the conformal limit, each individual deconstructing particle field, $\lambda_j$, decouples from the standard model.  In  processes mediated by unparticles or in which an unparticle is produced, the number of available intermediate or final state particles, $\lambda_j$, increases to compensate for the decoupling.  For the decay of an unparticle to standard model particles, the number of available final state particles does not increase to compensate for the decoupling.  Hence the unparticle does not decay to standard model particles \cite{Stephanov:2007ry}.  If unparticles are produced abundantly in early times, they could overclose the universe.
\par
In standard cosmology, Big Bang Nucleosynthesis (BBN) explains the abundance of the light elements.  For unparticles not to interfere with BBN, they must freeze out at a temperature, $T_*$, greater than the temperature of BBN, $1$ MeV.  If $T_*$ is too large, an overabundance of unparticles could have been produced in the early universe.  Conservatively, we assume $T_*=1$ MeV.
\par
We will now calculate the present day density of unparticles.  A similar calculation was previously done for Kaluza Klein gravitons in large extra-dimensions \cite{Hall:1999mk}.  First, the contribution from single flavor neutrino-neutrino annihilation is calculated.  The spin averaged matrix element for $\nu\nu\rightarrow \lambda_j$ is

\begin{eqnarray}
\overline{\sum}|\mathcal{M}|^2=\kappa^2_\nu\frac{F^2_j}{16}\frac{s^2}{\Lambda^{2d_{\mathcal{U}}}_{\mathcal{U}}},
\label{eq:neutannih}
\end{eqnarray}
where $s$ is the center of mass energy squared for the process.

\par
The evolution of the number density of the field $\lambda_j$ is described by the Boltzmann equation:

\begin{eqnarray}
\dot{n}^{(j)}+3n^{(j)}H&=&\int \frac{d^3\mathbf{p}_{\nu}}{(2\pi)^3 2|\mathbf{p}_{\nu}|}\frac{d^3\mathbf{p}_{\overline{\nu}}}{(2\pi)^3 2|\mathbf{p}_{\overline{\nu}}|}\frac{d^3\mathbf{p}_{j}}{(2\pi)^3 2\sqrt{\mathbf{p}^2_{j}+M^2_j}}\\
&&\times(2\pi)^4\delta^4(p_j-p_{\nu}-p_{\overline{\nu}})\overline{\sum}|\mathcal{M}|^2e^{-|\mathbf{p}_{\nu}|/T}e^{-|\mathbf{p}_{\overline{\nu}}|/T}
\end{eqnarray}
Integrating the Boltzmann equation, it is found that

\begin{eqnarray}
s\dot{Y}^{(j)}=\dot{n}^{(j)}+3n^{(j)}H=\kappa^2_{\nu}\frac{M^5_j T F^2_j}{512\pi^3 \Lambda^{2d_{\mathcal{U}}}_{\mathcal{U}}}\mathcal{K}_1 \bigg{(}\frac{M_j}{T}\bigg{)}, \label{eqn:BOLTZINT}
\end{eqnarray}
where $\mathcal{K}_1$ is a Bessel function of the second kind.  The conservation of entropy was used to rewrite the Boltzmann equation in terms of the variable $Y_j=n_j/s$, where $s$ is entropy density.  
\par
The temperature, $T$, of radiation and time, $t$, are related by  \cite{Kolb:1990}

\begin{eqnarray}
t=0.301g^{-1/2}_*\frac{M_{pl}}{T^2},
\end{eqnarray}
where $M_{pl}$ is the Planck mass.  Since we are interested in times after unparticles freeze out, the relativistic degrees of freedom are $g_*=10.75$.  Using that the entropy density is proportional to $T^3$, the present-day number density of $\lambda_j$ from single flavor neutrino annihilation is found to be

\begin{eqnarray}
n^{(j)}_o=1.1\times10^{-5}\kappa_{\nu}^2T^3_0M_{pl}\frac{M_jF^2_j}{\Lambda^{2d_{\mathcal{U}}}_{\mathcal{U}}}\int^{\infty}_{M_k/T_*}dx~x^3\mathcal{K}_1(x)
\end{eqnarray}
where $T_0=1.96$ K is the current temperature of neutrinos.

\par
The states $\lambda_j$ can also be produced by photon-photon annihilation.  For simplicity we assume the dimensionless coupling constants $\kappa_\nu$ and $\kappa_\gamma$ are equal and relabel them as $\kappa$. 
The spin averaged matrix element squared for $\gamma\gamma\rightarrow \lambda_j$ is

\begin{eqnarray}
\overline{\sum}|\mathcal{M}|^2=\kappa^2\frac{F^2_j}{2}\frac{s^2}{\Lambda^{2d_{\mathcal{U}}}_{\mathcal{U}}}.
\end{eqnarray}
There is also a symmetry factor of 1/2 from initial state photons.  Hence, the photon-photon annihilation contribution to the number density of tensor unparticles will be four times that of the single flavor neutrino-neutrino annihilation contribution.

\par
To find the present day density of tensor unparticles, it will be necessary to sum over all the states $\lambda_j$ and take the conformal limit.  From Eq.~(\ref{eq:massgap}), we see that summing over all states is equivalent to integrating the results for a single field $\lambda_j$ over the measure
\begin{eqnarray}
dn=dM^2_n/\Delta^2.
\label{eq:meas}
\end{eqnarray}
Considering all three flavors of neutrino, the present day density of unparticles from neutrino-neutrino and photon-photon annihilation is

\begin{eqnarray}
\rho_{\mathcal{U}}&=&\frac{14}{\Delta^2}\int^{\infty}_0dM_j~M^2_jn^{(j)}_0\\
&=&2.6\times10^{-5}\kappa^2A_{d_{\mathcal{U}}}M_{pl}T^3_0\bigg{(}\frac{T_*}{\Lambda_{\mathcal{U}}}\bigg{)}^{2d_{\mathcal{U}}}\int^{\infty}_0dy~y^{2d_{\mathcal{U}}-1}\int^{\infty}_ydx~x^3\mathcal{K}_1(x),
\end{eqnarray}
where we have substituted for $F_j$.  Note that since $F^2_j\propto \Delta^2$ and $dn\propto \Delta^{-2}$, the density of unparticles does not depend on the mass gap.  The results of the density of tensor unparticles for $d_{\mathcal{U}}=4/3,~3/2,~5/3,$ and $2$ are given in Table \ref{tab:dens} (a).

\begin{table*}[tb]
\begin{tabular}{|l|c|c|c|c|}
\hline\hline
$d_{\mathcal{U}}$ & $4/3$&$3/2$&$5/3$&$2$\\
\hline
(a) $\rho_{\mathcal{U}}$&$5.9\times10^{-39}\kappa^2\Lambda^{-8/3}_{\mathcal{U}}$&$6.1\times10^{-41}\kappa^2\Lambda^{-3}_{\mathcal{U}}$&$5.6\times10^{-43}\kappa^2\Lambda^{-10/3}_{\mathcal{U}}$&$3.7\times10^{-47}\kappa^2\Lambda^{-4}_{\mathcal{U}}$\\
\hline
(b) $\Lambda_{\mathcal{U}}>$&$1.9\times10^3\kappa^{3/4}$&$180\kappa^{2/3}$&$26\kappa^{3/5}$&$1.4\kappa^{1/2}$\\
\hline\hline
\end{tabular}
\caption{(a) The present day density of tensor unparticles from neutrino-neutrino and photon-photon annihilation in units of GeV$^4$, where $\Lambda_{\mathcal{U}}$ has been normalized to $1$ TeV, and (b) lower bounds on $\Lambda_{\mathcal{U}}$ in units of TeV for a variety $d_{\mathcal{U}}$ values.}
\label{tab:dens}
\end{table*}

\par

For tensor unparticles not to overclose the universe, their present day density must be less than the matter density of the universe, $\rho_m=1.028\times 10^{-47}$ GeV$^4$.  Using this constraint, bounds on $\Lambda_{\mathcal{U}}$ and $\kappa$ can be found.  These bounds are given in Table \ref{tab:dens} (b) for $d_{\mathcal{U}}=4/3,~3/2,~5/3,$ and $2$.

\par
The cosmology of scalar and vector unparticles has been studied before \cite{McDonald:2007bt,Davoudiasl:2007jr}.  We now compare these results to our results for tensor unparticles.
\par
In \cite{McDonald:2007bt}, the effects on BBN from quark decay to scalar unparticles was studied.    
It was found that in order for scalar unparticles not to interfere with BBN, $M_{\mathcal{U}}\gtrsim 20-2600$ TeV for $1.1\leq d_{\mathcal{U}}\leq 2$, $2\leq d_{\mathcal{BZ}}\leq 4$, $\Lambda_{\mathcal{U}}\gtrsim 1$ TeV, and  $A_{d_{\mathcal{U}}}C^2_{\mathcal{U}}\sim 1$.  Using these parameter values and the definition of $\kappa_{\mathcal{U}}$ in Eq.~(\ref{eqn:UNPARTOP}), for tensor unparticles we find the less stringent condition that $M_{\mathcal{U}}\gtrsim 1.8-180$ TeV for $4/3\le d_{\mathcal{U}}\le 2$.  The lowest bound came from $d_{\mathcal{U}}=2$ and $k=4$ and the highest bound came from $d_{\mathcal{U}}=4/3$ and $k=2$. 

\par
The effect of vector unparticles on BBN was studied in \cite{Davoudiasl:2007jr}.  With $k=2$, $M_{\mathcal{U}}=1000$ TeV, and $|C_{\mathcal{U}}|=1$, it was found that $\Lambda_{\mathcal{U}}\lesssim 100$ GeV for $d_{\mathcal{U}}=3/2$ and $2$.  There were no limits given for $d_{\mathcal{U}}=4/3$ or $5/3$.  For $k=d_{\mathcal{U}}=2$ there is no tensor unparticle density limit on $\Lambda_{\mathcal{U}}$, only a lower bound on $M_{\mathcal{U}}$.  For $d_{\mathcal{U}}=3/2$ we find that $\Lambda_{\mathcal{U}}\lesssim 1.7\times10^5$ TeV.  That is, under these assumptions we do not have a bound for $\Lambda_{\mathcal{U}}$.

\section{Supernovae Energy Loss Rates}
\label{sec:emissivity}
Data from IMB and Kamiokande indicates that in a period on the order of $10$ seconds the neutrino flux from SN1987A carried away more than $2\times 10^{53}$ ergs of energy.  The energy released due to the core collapse to a neutron star is $\sim 3\times 10^{53}$ ergs, hence neutrinos carry away most of this energy.  This places a constraint on the supernovae energy loss rate from new physics.  For a report on the classic example of supernovae energy losses through axions see \cite{Raffelt:1990yz}.  There have also been many studies of energy losses due to KK gravitons \cite{Barger:1999jf,Cullen:1999hc,Hanhart:2000er,Hanhart:2001fx,Hannestad:2001jv,Hannestad:2003yd,Fairbairn:2001ab,Kumar:2007gb}.  
\par
Since unparticles have a continuous mass spectrum, they can be produced in supernovae.  If the energy loss due to unparticles is excessive it can change stellar evolution.  Here we calculate the supernovae volume energy loss rates (emissivities) resulting from emission of tensor unparticles in photon-photon and electron-positron annihilation, and place constraints on $\Lambda_{\mathcal{U}}$ and the relevant dimensionless coupling constant.

\subsection{Photon-Photon annihilation}
\par
First we consider the photon-photon annihilation to tensors unparticles.  Using the Feynman rules derived in \cite{Cheung:2007ap}, we obtain the spin-averaged cross section for $\gamma\gamma\rightarrow\lambda_j$,

\begin{eqnarray}
\sigma^j_{\gamma\gamma}(s)=C^2_\gamma\frac{A_{d_{\mathcal{U}}}}{8}\frac{\Delta^2(M^2_j)^{d_{\mathcal{U}}-1}}{\Lambda^{2d_{\mathcal{U}}}_{\mathcal{U}}}\delta(s-M^2_j),
\end{eqnarray}
where $s$ is the squared center of mass energy.

\par
The supernova volume emissivity is found by thermally averaging over the Bose-Einstein distribution,
\begin{eqnarray}
Q^j_\gamma=&&\int\frac{2d^3{\bf k}_1}{(2\pi)^3}\frac{1}{e^{\omega_1/T}-1}\int\frac{2d^3{\bf k}_2}{(2\pi)^3}\frac{1}{e^{\omega_2/T}-1}\frac{s(\omega_1+\omega_2)}{2\omega_1\omega_2}\sigma^j_{\gamma\gamma},
\end{eqnarray}
where $T$ is the supernova temperature.
The center of mass energy squared in terms of the photon energies and opening angle, $\theta_{\gamma\gamma}$, is
\begin{eqnarray}
s=2\omega_1\omega_2(1-\cos\theta_{\gamma\gamma})
\end{eqnarray}
\par
To find the volume emissivity from tensor unparticles, we need to sum over the states $\lambda_j$ using the measure in Eq.~(\ref{eq:meas}).  After carrying out the integral and summing over states, the supernova volume emissivity from photon-photon annihilation to tensor unparticles is

\begin{eqnarray}
Q_{\gamma}=&&\zeta(d_{\mathcal{U}}+2)\Gamma(d_{\mathcal{U}}+2)\zeta(d_{\mathcal{U}}+3)\Gamma(d_{\mathcal{U}}+3)\kappa^2_\gamma\frac{A_{d_{\mathcal{U}}}4^{d_{\mathcal{U}}-1}}{2\pi^4(d_{\mathcal{U}}+1)}\frac{T^{2d_{\mathcal{U}}+5}}{\Lambda^{2d_{\mathcal{U}}}_{\mathcal{U}}}.
\end{eqnarray}
Plasma effects which may change the photon dispersion relations away from those of free particles \cite{Raffelt:1990yz} have been neglected.

\subsection{Electron Positron Annihilation}

In calculating the electron-positron annihilation cross section, we neglect the electron mass since $m_e<<T_{SN}\sim 30$ MeV, the benchmark supernova temperature.  The spin averaged cross section for electron positron annihilation to $\lambda_j$ is

\begin{eqnarray}
\sigma^j_{e^+e^-}(s)=\kappa^2_e\frac{A_{d_{\mathcal{U}}}}{16}\frac{\Delta^2(M^2_j)^{d_{\mathcal{U}}-1}}{\Lambda^{2d_{\mathcal{U}}}_{\mathcal{U}}}\delta(s-M^2_j)
\end{eqnarray}

\par

The supernova volume emissivity from the electron-positron annihilation to tensor unparticles is found by thermally averaging over the Fermi-Dirac distribution,

\begin{eqnarray}
Q^j_e=&&\int\frac{2d^2\vec{k}_1}{(2\pi)^3}\frac{1}{e^{(E_1-\mu_e)/T}+1}\int\frac{2d^3\vec{k}_2}{(2\pi)^3}\frac{1}{e^{(E_2+\mu_e)/T}+1}\frac{s(E_1+E_2)}{2E_1E_2}\sigma^j_{e^-e^+},
\end{eqnarray}
where $\mu_e$ and $-\mu_e$ are the electron and positron chemical potentials, respectively, and $T$ is the supernova temperature.  In the core of a supernova $\mu_e\simeq(3\pi^2n_e)^{1/3}\simeq 345$ MeV.  After performing the integration and summing over possible final states, we find the volume emissivity from electron-positron annihilation to tensor unparticles is

\begin{eqnarray}
Q_e=\kappa^2_e\frac{4^{d_{\mathcal{U}}-1}A_{d_{\mathcal{U}}}I_e(d_{\mathcal{U}})}{8\pi^4(d_{\mathcal{U}}+1)}\frac{T^{2d_{\mathcal{U}}+5}}{\Lambda^{2d_{\mathcal{U}}}_{\mathcal{U}}},
\end{eqnarray}
where

\begin{eqnarray}
I_e(d_{\mathcal{U}})=\int dy_1dy_2\frac{(y_1y_2)^{d_{\mathcal{U}}+1}(y_1+y_2)}{(e^{y_1-\mu_e/T}+1)(e^{y_2+\mu_e/T}+1)}.
\end{eqnarray}
For a supernova temperature of $30$ MeV, the values of $I_e(d_{\mathcal{U}})$ are $0.41$ for $d_{\mathcal{U}}=4/3$, $0.73$ for $d_{\mathcal{U}}=3/2$, $1.3$ for $d_{\mathcal{U}}=5/3$, and $4.3$ for $d_{\mathcal{U}}=2$.

\subsection{Limits on $\Lambda_{\mathcal{U}}$ and $\kappa$}
\par
\begin{table*}[t]
\begin{tabular}{|l|c|c|c|c|}
\hline\hline
$d_{\mathcal{U}}$& $4/3$&$3/2$&$5/3$&$2$\\
\hline
(a) $Q_{\gamma}$&$3.6\times 10^{41}\kappa^2_\gamma\Lambda^{-8/3}_{\mathcal{U}}$&$1.2\times10^{40}\kappa^2_\gamma\Lambda^{-3}_{\mathcal{U}}$&$73.8\times10^{38}\kappa^2_\gamma\Lambda^{-10/3}_{\mathcal{U}}$&$2.7\times10^{35}\kappa^2_\gamma\Lambda^{-4}_{\mathcal{U}}$\\
\hline
(b) $Q_{e}$&$1.2\times10^{39}\kappa^2_e\Lambda^{-8/3}_{\mathcal{U}}$&$4.9\times10^{37}\kappa^2_e\Lambda^{-3}_{\mathcal{U}}$&$1.8\times10^{36}\kappa^2_e\Lambda^{-10/3}_{\mathcal{U}}$&$1.8\times10^{33}\kappa^2_e\Lambda^{-4}_{\mathcal{U}}$\\
\hline
\hline
(c) $\Lambda_{\mathcal{U}}:\gamma\gamma$ & $1.0\times10^3~\kappa^{3/4}_{\gamma}$ &$160~\kappa^{2/3}_\gamma$& $33~\kappa^{3/5}_{\gamma}$ & $3.1~\kappa^{1/2}_{\gamma}$\\
\hline
(d) $\Lambda_{\mathcal{U}}:e^-e^+$ &$125~\kappa^{3/4}_{e}$&$25~\kappa^{2/3}_e$&$6.8~\kappa^{3/5}_{e}$&$0.88~\kappa^{1/2}_e$\\
\hline\hline
\end{tabular}
\caption{Volume emissivities of (a) photon-photon and (b) electron-positron annihilation in units of erg cm$^{-3}$ s$^{-1}$, where $\Lambda_{\mathcal{U}}$ has been normalized to $1$ TeV, and lower bounds on $\Lambda_{\mathcal{U}}$ in units of TeV from SN1987A for (c) photon-photon and (d) electron-positron annihilation for various values of $d_{\mathcal{U}}$ and a supernova temperature of $30$ MeV.}
\label{tab:em}
\end{table*}

To prevent the neutrino burst from SN1987A from being too short, the upper limit on the supernova volume emissivity of new physics is \cite{Raffelt:1990yz,Raffelt:1999tx}

\begin{eqnarray}
Q_{SN}\sim 3\times10^{33}\text{ erg cm}^{-3}\text{ s}^{-1}.
\end{eqnarray}
\par
Using this upper bound on the volume emissivities from photon-photon and electron-positron annihilation to tensor unparticles, we obtain limits on $\Lambda_{\mathcal{U}}$ and $\kappa$ for a supernova temperature of $30$ MeV.  The results for the volume emissivities and bounds for $d_{\mathcal{U}}=4/3,~3/2,~5/3,$ and $2$ are presented in Table \ref{tab:em}.
The bounds from photon-photon annihilation are on the same order as those from the matter density, while the bounds from electron-positron annihilation are an order of magnitude less.
\par
Previous studies have calculated the volume emissivity bounds from nucleon brehmstrahllung for vector unparticles \cite{Freitas:2007ip, Hannestad:2007ys,Davoudiasl:2007jr} and scalar unparticles \cite{Freitas:2007ip}.  The constraints on vector and scalar unparticle couplings to the standard model from 5th force experiments were also calculated in \cite{Freitas:2007ip}.  
\par
The operators under study in Ref.~\cite{Freitas:2007ip} had coefficients $\kappa/M^{d_{\mathcal{U}}+d_{SM}-4}_Z$ instead of $\kappa/\Lambda^{d_{\mathcal{U}}+d_{SM}-4}_{\mathcal{U}}$.  Hence, to compare our results we replace $\Lambda_{\mathcal{U}}$ with $M_Z$ and find the constraints on the dimensionless coupling constant, $\kappa$.   Table \ref{tab:comparison} contains the bounds from supernovae volume emissivity and 5th force experiments on parity conserving scalar and vector unparticle and standard model interactions \cite{Freitas:2007ip}, and our bounds on tensor unparticle and standard model interactions.  
\par
In \cite{Hannestad:2007ys} it was found that for a vector unparticle coupling to nucleons, data from SN1987A gave the bounds
\begin{eqnarray}
\frac{\kappa}{\Lambda^{d_{\mathcal{U}}-1}_{\mathcal{U}}}(30~\text{MeV})^{d_{\mathcal{U}}-1}=C_{\mathcal{U}}\frac{\Lambda^{d_{\mathcal{BZ}}-d_{\mathcal{U}}}_{\mathcal{U}}}{M^{d_{\mathcal{BZ}}-1}_{\mathcal{U}}}(30~\text{MeV})^{d_{\mathcal{U}}-1}\lesssim3\times10^{-11},
\end{eqnarray}
where $d_{SM}=3$.  Again, to compare to the results in \cite{Freitas:2007ip} we replace $\Lambda_{\mathcal{U}}$ with $M_Z$.  These results are presented in Table \ref{tab:comparison}.

\begin{table*}[tb]
\begin{tabular}{@{\extracolsep{1pt}}|l|c|c|c|c|}
\hline
$d_{\mathcal{U}}$&$4/3$&$3/2$&$5/3$&$2$\\
\hline
\multicolumn{5}{|c|}{Tensor Unparticles}\\
\hline
Unparticle Density &$1.7\times10^{-6}$&$1.1\times10^{-5}$&$7.9\times10^{-5}$&$4.4\times10^{-3}$\\
\hline
SN1987A:$\gamma\gamma$    &$3.8\times10^{-6}$&$1.4\times10^{-5}$&$5.2\times10^{-5}$&$8.7\times10^{-4}$\\
\hline
SN1987A:$e^-e^+$    &$6.6\times10^{-5}$&$2.2\times10^{-4}$&$7.6\times10^{-4}$&$1.1\times10^{-2}$\\
\hline
\multicolumn{5}{|c|}{Vector Unparticles}\\
\hline
5th force\cite{Freitas:2007ip}    &$1.4\times10^{-15}$&--&$1.8\times10^{-10}$&$2\times10^{-5}$\\
\hline
SN1987A\cite{Freitas:2007ip}    &$3.5\times10^{-8}$&--&$1\times10^{-6}$&$3\times10^{-5}$\\
\hline
SN1987A\cite{Hannestad:2007ys}    &$4.3\times10^{-10}$&$1.7\times10^{-9}$&$6.3\times10^{-9}$&$9.1\times10^{-8}$\\
\hline
\multicolumn{5}{|c|}{Scalar Unparticles}\\
\hline
5th force\cite{Freitas:2007ip}    &$1.2\times10^{-13}$&--&$1.6\times10^{-8}$&$1.7\times10^{-3}$\\
\hline
SN1987A\cite{Freitas:2007ip}    &$2.4\times10^{-6}$&--&$6.6\times10^{-5}$&$2\times10^{-3}$\\
\hline

\end{tabular}
\caption{Comparison of upper bounds on the coupling constant $\kappa$ from tensor unparticle density, supernovae volume energy loss rates for scalar \cite{Freitas:2007ip}, vector \cite{Freitas:2007ip,Hannestad:2007ys}, and tensor unparticles, and fifth force experiments for vector and scalar unparticles \cite{Freitas:2007ip}. The conformal scale $\Lambda_{\mathcal{U}}$ is taken to be $M_Z$.  Entries with dashes are not available}
\label{tab:comparison}
\end{table*}
\par
As can be seen in Table \ref{tab:comparison},  our bounds for tensor unparticle and standard model couplings are weaker than those for vector unparticle and standard model couplings \cite{Freitas:2007ip,Hannestad:2007ys}.  Our results are weaker than the fifth force experiment bounds on scalar unparticle couplings for $d_{\mathcal{U}}=4/3$ and $5/3$, and on the same order of magnitude for $d_{\mathcal{U}}=2$ \cite{Freitas:2007ip}.  The tensor unparticle bounds are on the same order of magnitude as the supernovae volume emissivity bounds on scalar unparticle couplings \cite{Freitas:2007ip}.

\section{Conclusion}
\label{sec:Conclusion}
We calculated the present day density of tensor unparticles from photon-photon and neutrino-neutrino annihilation and the supernovae volume energy loss rates from photon-photon and electron-positron annihilation to tensor unparticles.  The deconstruction of the unparticle given in \cite{Stephanov:2007ry} was used to calculate these observables. 
\par
Using the matter density of the universe as an upper bound on the present day tensor unparticle density, we placed constraints on the conformal scale $\Lambda_{\mathcal{U}}$ and the dimensionless coupling constant $\kappa_\gamma=\kappa_\nu=\kappa$. The supernovae volume emissivity bounds on new physics from SN1987A were used to place bounds on $\Lambda_{\mathcal{U}}$ and the individual coupling constants between tensor unparticles and photons, $\kappa_\gamma$, and tensor unparticles and electrons, $\kappa_e$.  
\par
Bounds for supernovae volume emissivities from electron-positron annihilation were an order of magnitude less than those from tensor unparticle density and supernovae volume emissivity from photon-photon annihilation to tensor unparticles.  
The bounds on the dimensionless coupling constants and $\Lambda_{\mathcal{U}}$ decreased as $d_{\mathcal{U}}$ increased.
Taking $\Lambda_{\mathcal{U}}=1$ TeV and assuming that all the dimensionless coupling constants are equal, the most stringent bounds we found were
\begin{eqnarray}
\kappa<\left\{\begin{array}{rl}
	4.3\times10^{-5}&\mbox{ for $d_{\mathcal{U}}=4/3$}\\
	4.1\times10^{-4}&\mbox{ for $d_{\mathcal{U}}=3/2$}\\
	2.9\times10^{-4}&\mbox{ for $d_{\mathcal{U}}=5/3$}\\
	1.0\times10^{-1}&\mbox{ for $d_{\mathcal{U}}=2$}
		\end{array} \right.
\end{eqnarray}
These bounds are weaker than those found from supernovae volume emissivities \cite{Freitas:2007ip,Hannestad:2007ys} and fifth force experiments \cite{Freitas:2007ip} for vector unparticles.  For $d_{\mathcal{U}}<2$, the constraints for tensor unparticles are weaker than the 5th force bounds for scalar unparticles, but for $d_{\mathcal{U}}=2$ they are on the same order of magnitude \cite{Freitas:2007ip}.  Our bounds for tensor unparticles are on the same order of magnitude as the supernovae volume energy loss rate bounds for scalar unparticles \cite{Freitas:2007ip}.
\par
The collider phenomenology of tensor unparticles at the LHC and electron positron-colliders has been studied previously \cite{Cheung:2007ap,Majhi:2007tu,Alan:2007ss}.  Their results for tensor unparticles at electron positron and hadron colliders were found for $\Lambda_{\mathcal{U}}\le1$ TeV and $\kappa=1$.  Our results from cosmological and astrophysical constraints show that the value of $\kappa$ is bounded by at least one order of magnitude lower for $d_{\mathcal{U}}=2$ and at least 5 orders of magnitude lower for $d_{\mathcal{U}}=4/3$, making signals at future colliders more difficult to observe.
\\
\\
\\
{\bf Acknowledgements}\\
I would like to thank Prof.~Tao Han for suggesting this project,
guiding it through, and commenting on the manuscript.
This work was supported in part by the U.S.~Department of Energy
under grant DE-FG02-95ER40896 and in part by the Wisconsin
Alumni Research Foundation.

%%%%%%%%%%%%%%%
\bibliographystyle{apsrev}
\bibliography{unpartens}                      %name of .bib file
%%%%%%%%%%%%%%%s

\end{document}